\documentclass[10pt]{article}
\usepackage{amssymb,latexsym}
\usepackage{amsmath}
\usepackage{amsfonts}
\usepackage{amsthm}
\usepackage{graphicx}
\usepackage{subfigure}
\usepackage[import,arrow,line,curve]{xy}

\theoremstyle{definition}

\makeatletter
\def\keywords{
    \vspace{1ex}
    \noindent
    \if@twocolumn
      \small{\bf  Keywords}\/---$\!$    \else
      \begin{center}
      \small\ {\bf Keywords}
      \end{center}
      \quotation\small
    \fi}
\def\endkeywords{\vspace{0.6em}\par\if@twocolumn\else\endquotation\fi
    \normalsize\rm}
\def\keywrds{
    \vspace{1ex}
    \noindent
    \if@twocolumn
      \small{\bf Mathematics subject classifications}\/---$\!$    \else
      \begin{center}\small\ {\bf AMS subject
classifications}\end{center}\quotation\small
    \fi}
\def\endkeywrds{\vspace{0.6em}\par\if@twocolumn\else\endquotation\fi
    \normalsize\rm}
\makeatother

\begin{document}
\title{On kink-dynamics of Stacked-Josephson Junctions}
\author{H. Susanto, T.P.P Visser, \& S. A. van Gils \\
{\small Department of Applied Mathematics,}\\
{\small University of Twente, P.O. Box 217, 7500 AE Enschede,}\\
{\small The Netherlands}
}
\date{}
\maketitle
\begin{abstract}
Dynamics of a fluxon in a stack of coupled long Josephson junctions is
studied numerically. Based on the numerical simulations, we show that
the dependence of the propagation velocity $c$ on the external
bias current $\gamma$ is determined by the ratio of the critical
currents of the two junctions $J$.
\end{abstract}

\section{Introduction}

The stacked-Josephson junctions we consider here consist of three
slabs of superconducting material which have an insulating barrier
between two superconductors. An important application of these
long Josephson junctions (LJJ) is the magnetic flux quanta
(fluxons). The fluxon, which is a circulating current, can move
along the junctions if there is a biased current applied.

The one dimensional stacked-LJJ can be modelled by a perturbed
sine-Gordon equation which, in normalized form, may be written as
\cite{mms}:
\begin{equation}
\begin{array}{ll}
\phi^{1}_{xx}-\phi^{1}_{tt}-
\sin{\phi^{1}}-S\phi^{2}_{xx}=\alpha\phi^{1}_t-\gamma, \\
\phi^{2}_{xx}-\phi^{2}_{tt}-
\sin{\phi^{2}}/J-S\phi^{1}_{xx}=\alpha\phi^{2}_t-\gamma.
\end{array}
\label{cjjs}
\end{equation}
Here $\phi^i$ is the superconducting phase difference across the
junction $i$ \cite{bp}. The term $\alpha\phi_t$ models the damping
due to quasi particle tunnelling, $\alpha>0$. The driving force,
corresponding to the normalized bias current $\gamma$, typically
lies in the interval $0\leq\gamma\leq 1$. $S\in [-1,0]$ is the
coupling parameter of the two junctions and $J$ is the ratio of
the critical current of of the two junctions.

One of the solutions of the unperturbed-sine-Gordon equation
($\phi_{xx}-\phi_{tt}=\sin\phi$) is the soliton
\begin{equation}
\phi_0=4\arctan{[ \exp{(\frac{x-ct-\xi_0}{\sqrt{1-c^2}})}]},
\end{equation}
which is moving with velocity $c$, $|c|<1$. In the unperturbed
equation this parameter is not determined. The wave speed will
become a function of $\gamma$ when $\alpha$ and $S$ do not vanish
and the damping is balanced by the driving force. Therefore, $c$
is a function of $\gamma$ for given $\alpha$ and $S$. The free
parameter $\xi_0$ determines the initial position of the fluxon.
In physics, this soliton solution is related with units of the
flux quantum or fluxon. The system (\ref{cjjs}) has, when
$S=\alpha=\gamma=0$, the solution $(\phi_0,0)$ in the primary
state $[1|0]$.

In this paper, we investigate the relation of $c$ and $\gamma$ for
spesific values of $\alpha$ and $S$ in the primary state $[1|0]$.
To represent the configuration, the notation $[n|m]$ is used to
denote the state where $n$ solitons are located in the first
junction and $m$ solitons in the other. Negative numbers are used
for antisolitons. In Sec.~\ref{sec1} we consider the case of
identical junctions. Unequal junctions are considered in
Secs.~\ref{sec2} and \ref{sec3}. The junctions considered in these
sections differ only on the critical currents. In Sec.~\ref{sec4}
we draw conclusions.

\section{Identical Junctions}
\label{sec1}

We look at travelling wave solutions to the equation (\ref{cjjs}).
So we assume that the solution only depends on $\xi=x-ct$. The
partial differential equation is thus reduced to a system of
ordinary differential equations
\begin{equation}
\begin{array}{ll}
(1-c^2)\phi^{1}_{\xi\xi}-
\sin{\phi^{1}}-S\phi^{2}_{\xi\xi}=-c\alpha\phi^{1}_\xi-\gamma, \\
(1-c^2)\phi^{2}_{\xi\xi}-
\sin{\phi^{2}}/J-S\phi^{1}_{\xi\xi}=-c\alpha\phi^{2}_\xi-\gamma.
\end{array}
\label{sjj}
\end{equation}

\begin{figure}[h]
\begin{center}
\subfigure[]{\includegraphics[width=7cm,angle=-0]{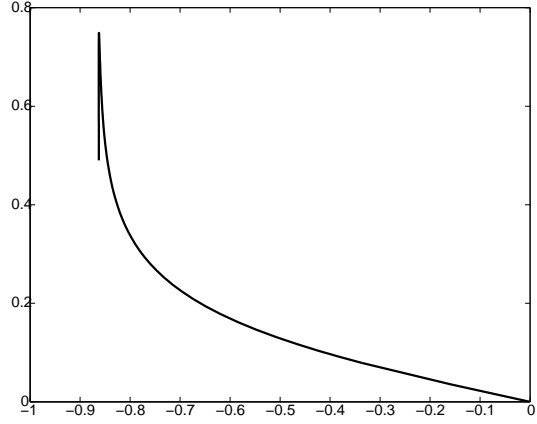}}
\hspace{0.5cm}
\subfigure[]{\includegraphics[width=7cm,angle=-0]{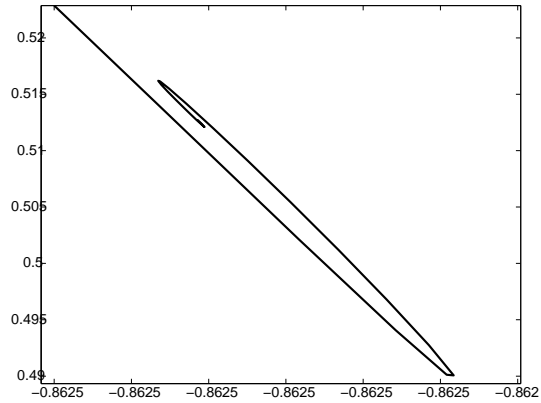}}
\end{center}
\caption{The relation between $\gamma$ (vertical axis) and $c$
(horizontal axis) which is known as the IV-characteristic is presented. Picture (a) shows the complete relation of $\gamma$ and $c$. Picture (b) magnifies a part of the plot (a) where it becomes clear that the curve indeed contains a spiral.} \label{spicurve}
\end{figure}

Considering (\ref{sjj}) in $(\phi^1,\phi^1_\xi,\phi^2,\phi^2_\xi)^T$
phase space, the problem of computing solitary wave profiles and
speeds for (\ref{cjjs}) with $[1|0]$ state is equivalent to finding
heteroclinic orbits connecting two fixed points:
\begin{equation}
(\phi^1,\phi^2)=(\arcsin\gamma,0)
\end{equation}
and
\begin{equation}
(\phi^1,\phi^2)=(\arcsin\gamma+2\pi,0).
\end{equation}
We can calculate the parameter combinations for which there exist
such a heteroclinic solution numerically, using the program AUTO
\cite{auto}. First, we consider the case when $J=1$. In this case,
the system of equations (\ref{sjj}) can be rewritten to
\begin{equation}
\left[
(1-c^2)-\frac{S}{1-c^2}\right]\phi^{1,2}_{\xi\xi}-\sin\phi^{1,2}
-\frac{S}{1-c^2}\sin\phi^{2,1}=\frac{S}{1-c^2}(\alpha
c\phi^{2,1}_\xi-\gamma)+\alpha c\phi^{1,2}_\xi-\gamma
\end{equation}
from which we see a singularity at $1-c^2=\pm S$ or $|c_\pm
|=\sqrt{1\pm S}$. These values of $c$ are called the {\it Swihart
velocity}. Note that $|c_+|<|c_-|$ because $S$ is negative.

In Fig. \ref{spicurve}, the numerically obtained $(c,\gamma)$
picture (in the future we call these kind of pictures
IV-characteristics) is shown. The numerical values used in all
calculations are $S=-0.25$ and $\alpha=0.17$.

\begin{figure}[h]
\begin{center}
\subfigure[]{\includegraphics[width=7cm,angle=-0]{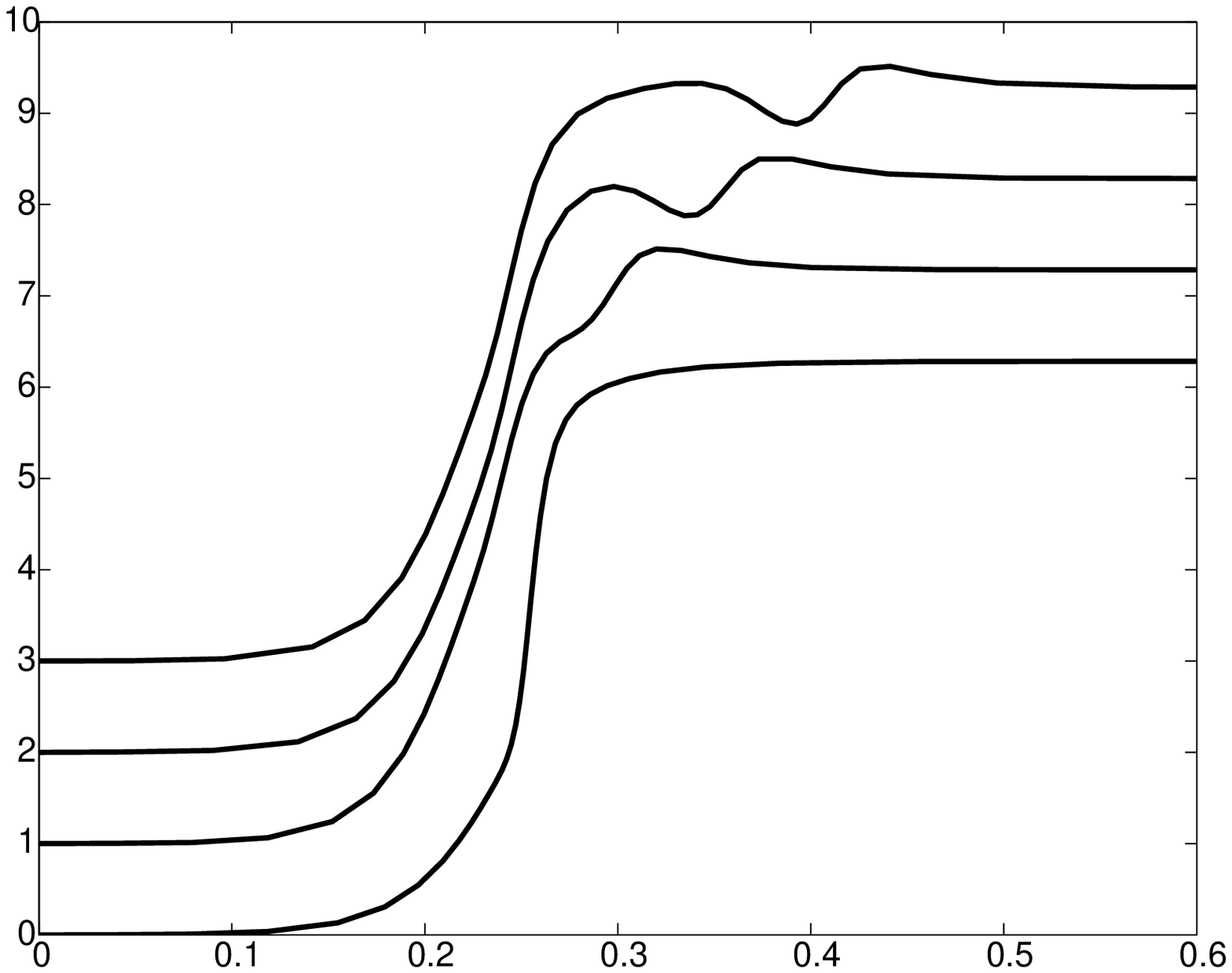}}
\hspace{0.5cm}
\subfigure[]{\includegraphics[width=7cm,angle=-0]{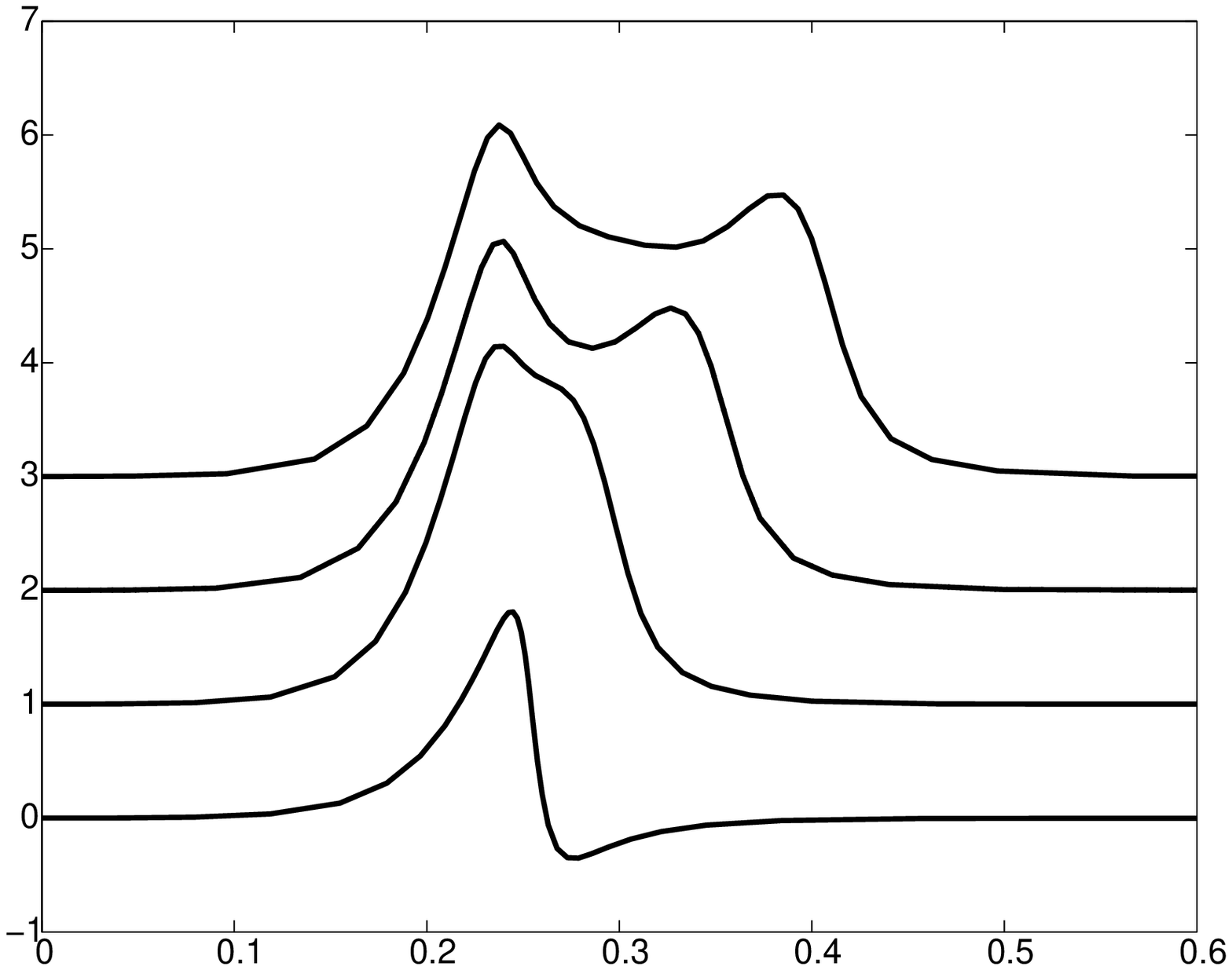}}
\end{center}
\caption{$\phi^1$ (a) and $\phi^2$ (b) as functions of $\xi$ at
the first four turning points of the spiral in the IV-characteristic. The lowest solution corresponds to the first turning point. The $i$-th solution is shifted $(i-1)$ unit(s) in the vertical direction for clarity. The space variable $x$ has been scaled such that it is in the interval $[0,1]$.} \label{soltosp}
\end{figure}

Somehow, near the edge of the IV-characteristic, we have a spiral
shaped curve. Previously, this kind of spiral in the
IV-characteristic has been observed in a single long Josephson
junction by Brown {\it et al.} \cite{dlb}. Later on, it is shown
in \cite{bgv,tv} that such a spiral exists in the single Josephson
junction if the dissipation due to surface resistance, {\it i.e.}
adding a term $\beta\phi^{1,2}_{xxt}$ in Eq.~(\ref{cjjs}), is
taken into account.

However, there is a difference between the spiral we have in this
case and spiral in a single long Josephson junction. The spiral in
the single long Josephson junction has its center located at
$|c|=1$ which is the maksimum velocity possibly achieved by a
fluxon. In our case here, the center is below the lowest Swihart
velocity $|c_+|=\sqrt{1+S}$ \cite{ukc}. The first four turning
points are given in Table \ref{tp1}. The plot of the solitons
related to those turning points is given in Figure \ref{soltosp}.

\begin{table}[ht]
\begin{center}
\begin{tabular}{c|c}
\hline
$\gamma$ & c \\ \hline
0.7489735 & -0.8622143\\ \hline
0.4900705 & -0.8625241\\ \hline
0.5161983 & -0.8625433\\ \hline
0.5121004 & -0.8625403\\ \hline
\end{tabular}
\end{center}
\caption{The first four turning points of the curve in the $\gamma-c$ space.}
\label{tp1}
\end{table}

Goldobin {\it et al.} \cite{gmu} give a 'proof' which shows that
the fluxon in the state $[1|0]$ with $J=1$ cannot pass the lowest
Swihart velocity $|c_+|$. The result says that if $|c_+|$ is
reached, there is at least a fluxon in the second junction which
then gives the $[1|1]$ state.

\section{Unequal Junctions: $J>1$}
\label{sec2}

An interesting situation occurs if the two junctions forming the
stack are not identical. We assume that only the critical currents
of the two are unequal, which translates to the  condition that
$J$ in Eq.~(\ref{sjj}) is not equal to $1$.

\begin{figure}[tbh]
\begin{center}
{\includegraphics[width=7cm,angle=-0]{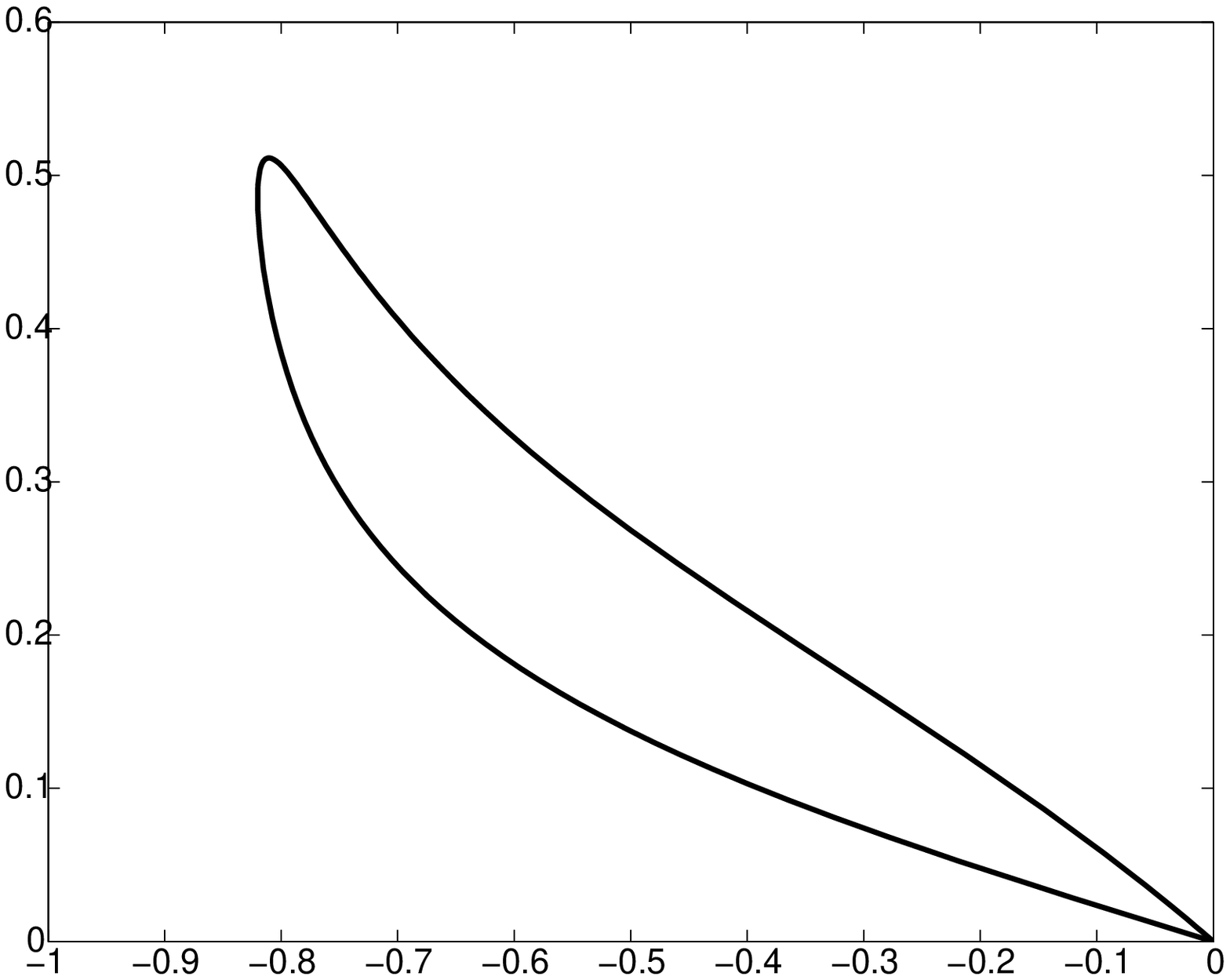}}
\end{center}
\caption{The IV-characteristic of the stacked Josephson junction with $J>1$. The critical current ratio $J$ in this calculation is taken to be $1.5$. Using AUTO, the 'back-bending' curve can be traced. It is known that solutions which lie on the upper/right branch are unstable.}
\label{backban}
\end{figure}

Goldobin {\it et al.} \cite{gwu}  mention for the first time that
there is a 'back-bending' phenomenon in the IV-characteristic. With
AUTO we can continue the backbending such that we get the
full-branch of the IV-characteristic. The result is shown in Fig.
\ref{backban}.

Goldobin {\it et al.} \cite{gmu} also show that the velocity
$|c_+|$ cannot be reached at this case.

\begin{figure}[h]
\begin{center}
\subfigure[]{\includegraphics[width=7cm,angle=-0]{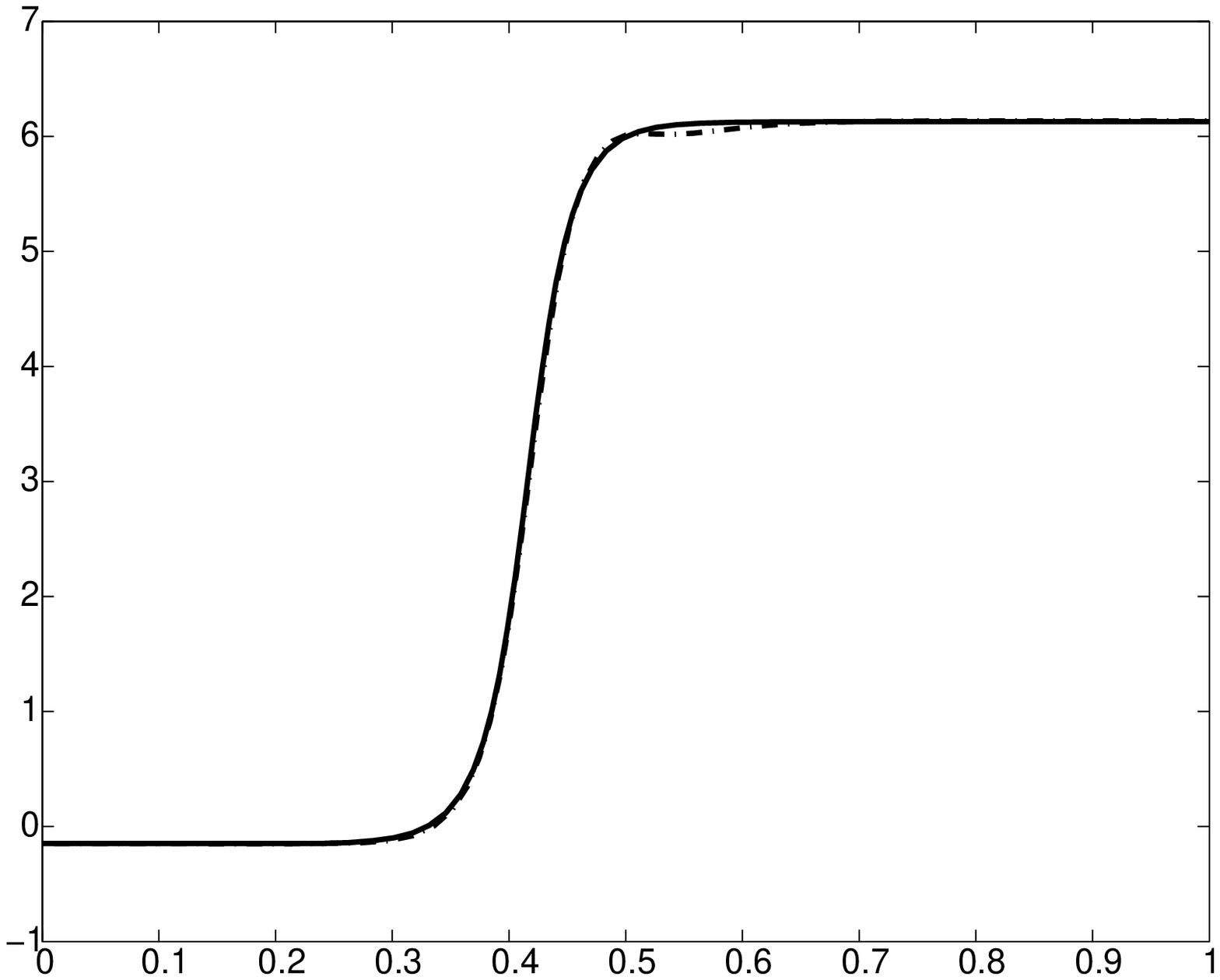}}
\hspace{0.5cm}
\subfigure[]{\includegraphics[width=7cm,angle=-0]{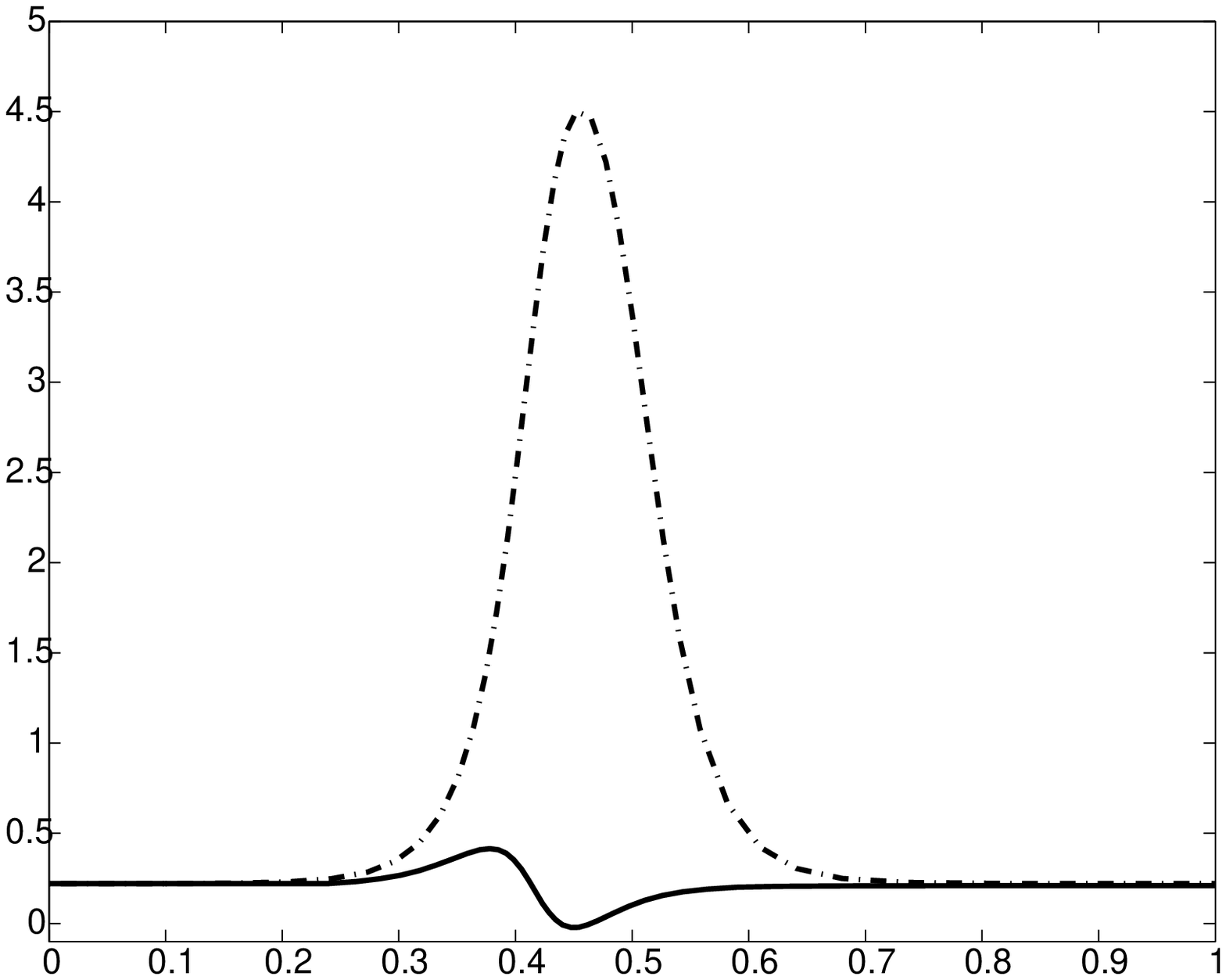}}
\end{center}
\caption{Two pairs of solution to (\ref{sjj}) for $J=1.5$ and $\gamma=0.17$ on the left and right branch of the IV-characteristic are compared. Picture (a) and (b) compare the $\phi^1$'s and $\phi^2$'s respectively. The solutions on the right branch is given in dashed-dotted line.}\label{solbackban}
\end{figure}

In Fig.~\ref{solbackban}, two solutions for the same value of
$\gamma$ are plotted. In the first junction there is hardly any
change, but in the second junction there is a big difference. On
the left branch there is only a small 'image' in the second
junction, which then if we follow the IV-characteristic, this
image has grown to a large hump.

The stability of solutions on the right branch has been tested
numerically in \cite{tv}. By taking the solution as calculated
with AUTO and using this as initial solution to the PDE-solver
\cite{tv1}, we conclude that all solutions on the right branch are
unstable. All solutions at the left branch up to the point where
$d\gamma/dc=0$ are stable. Solutions on the right branch are in
the domain of attraction of the solution for the same value of
$\gamma$ on the left branch.

The instability of the solution can be seen clearer when we follow
the right branch in the direction of decreasing $\gamma$. In the
limit $\gamma$ tends to $0$, there is a fluxon-antifluxon pair in
junction $2$ or the stack is in $[1|1,-1]$ \cite{gmu}. Because the
fluxon and antifluxon are attracting each other, if we apply a
bias current, they will collide and a new different state is
formed. This is the physical argument to show that this state is
unstable.

\section{Unequal Junctions: $J<1$}
\label{sec3}

A somewhat different behavior happens when we look at case the
$J<1$. While in the case $J\geq 1$, the velocity is bounded by
$c_+$, $|c|\leq |c_+|$, in the case $J<1$ we have that $|c|$ can
exceed the value $|c_+|$. When fluxons move with velocities above
the lowest Swihart velocity, {\it Cherenkov radiation} takes
place.

\begin{figure}[tbh]
\begin{center}
\subfigure[]{\includegraphics[width=7cm,angle=-0]{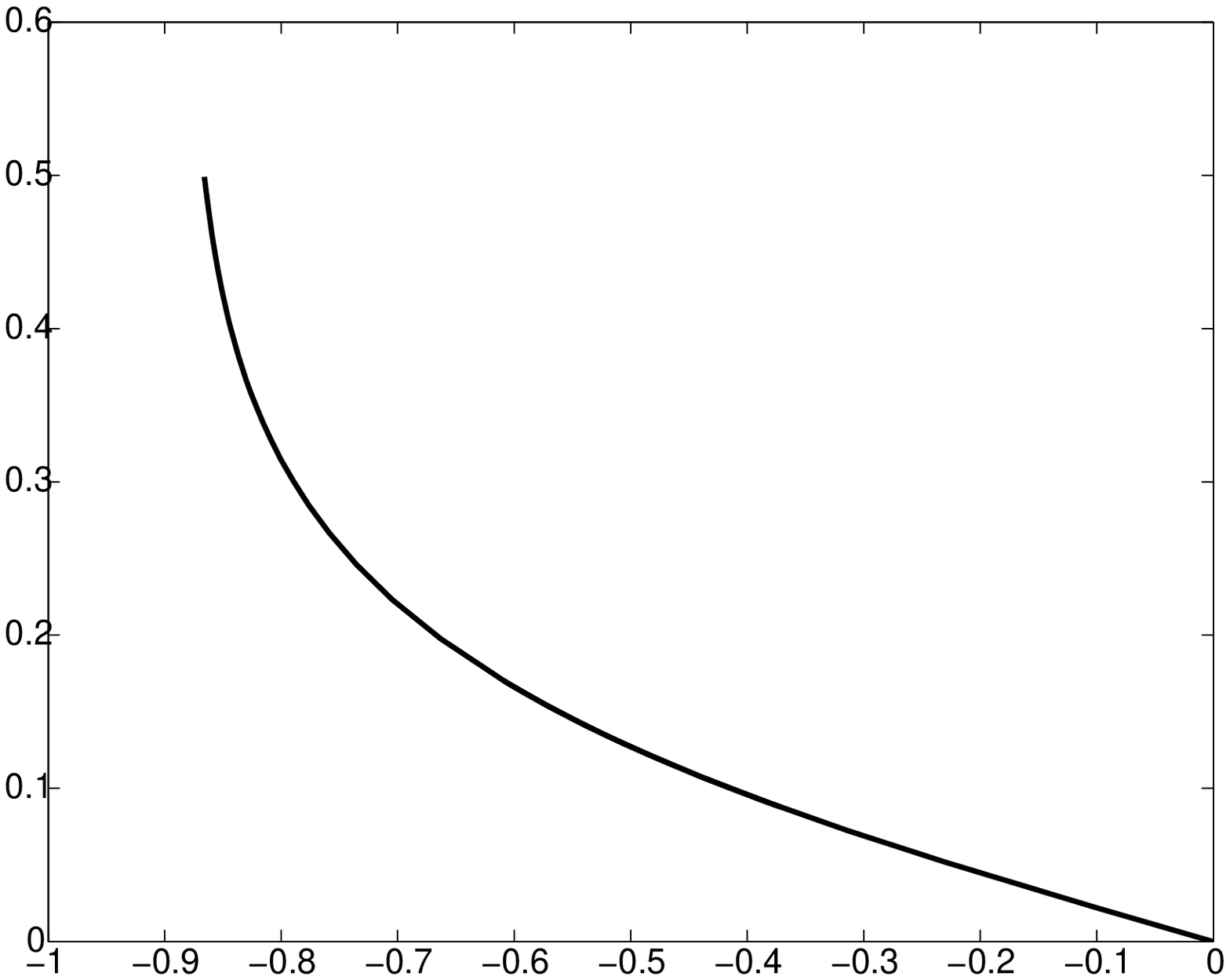}}
\hspace{0.5cm}
\subfigure[]{\includegraphics[width=7cm,angle=-0]{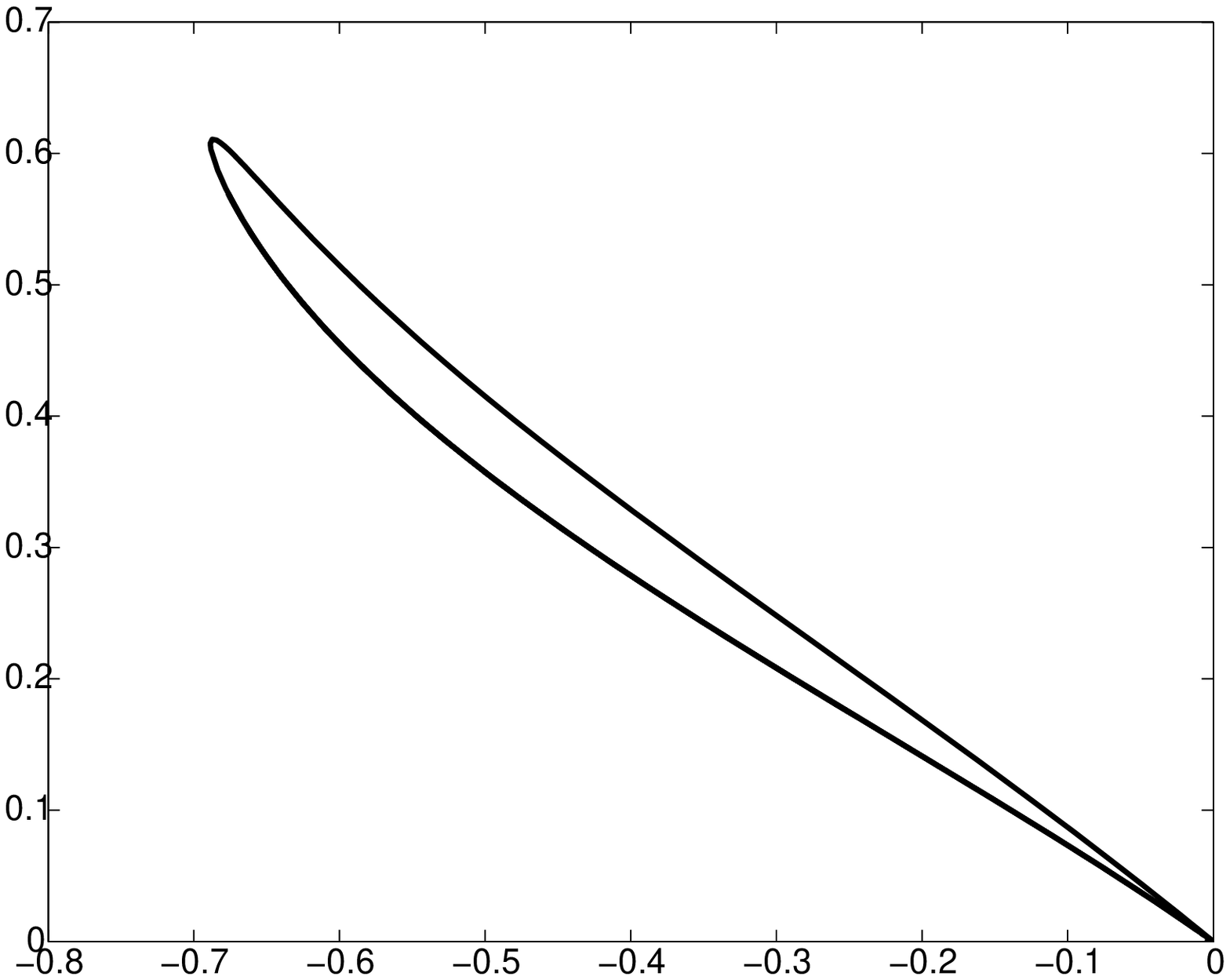}}
\end{center}
\caption{The same as Figure \ref{backban} but with $J<1$ where $J$ is taken to be $0.5$. In (a) we present the well known IV-characteristic where the curve can pass the lowest Swihart velocity $c_+$. Because of the singularity, the AUTO calculation here stops at $c_+$. In (b) another IV-characteristic is shown. All solutions lying on this 'back-bending' curve are unstable.}
\label{ivj}
\end{figure}

Cherenkov radiation is the emission of waves behind the moving
fluxon. Behind the fluxon we will find an oscillating tail which
is the emitted waves. E. Goldobin {\it et al.} \cite{gwu,gwtu}
observe Cherenkov radiation happening in the $[1|0]$ state
numerically and experimentally.

The IV-characteristic of the state is given in Fig.~\ref{ivj}(a).
Unfortunately, the IV-characteristic stops at $c_+$ because our
equation becomes singular at that value and AUTO cannot continue
the calculation through the singularity. Therefore, all solutions
that lie in this IV-characteristic do not have such emitted wave
behind the fluxon.

\begin{figure}[tbh]
\begin{center}
\subfigure[]{\includegraphics[width=7cm,angle=-0]{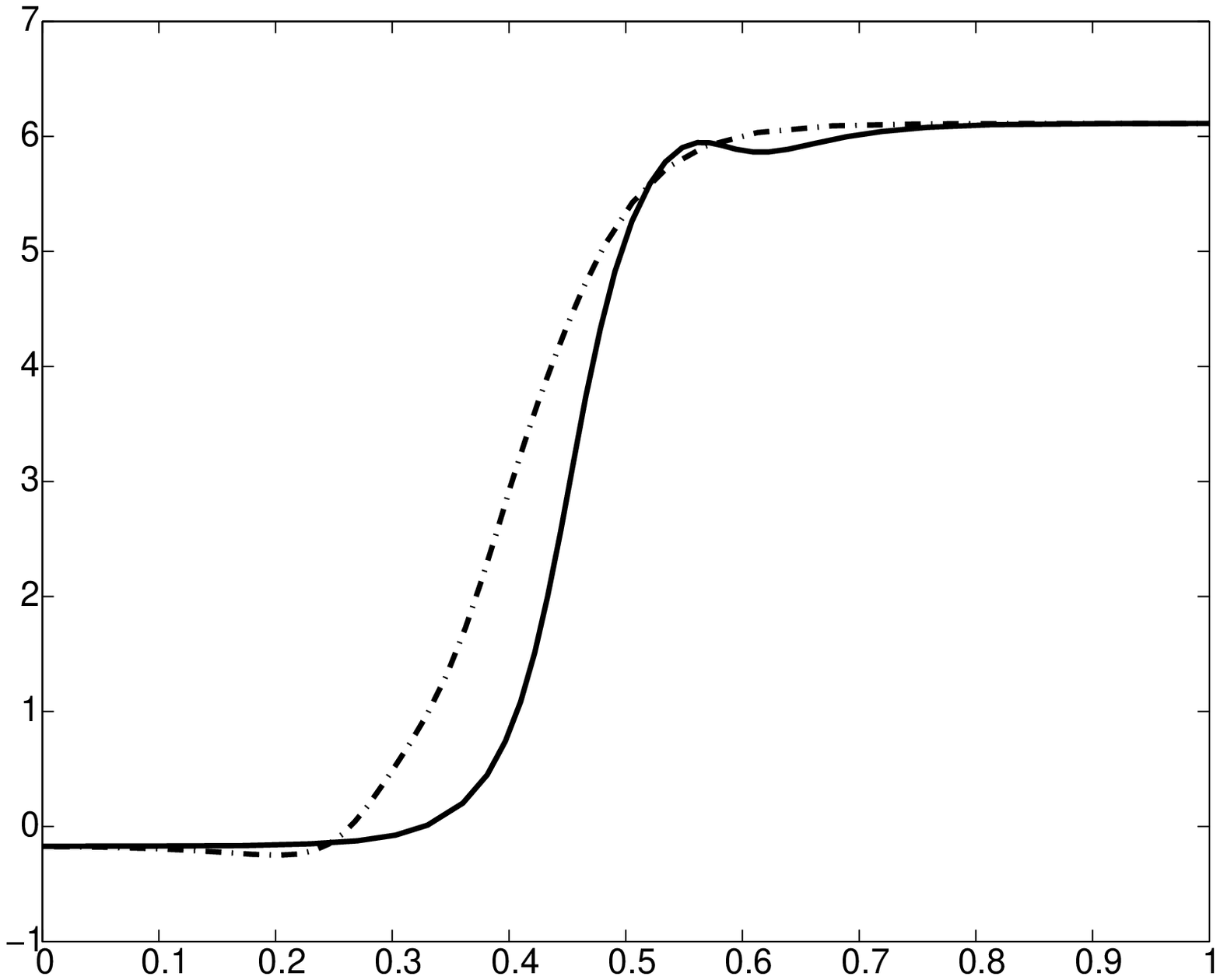}}
\hspace{0.5cm}
\subfigure[]{\includegraphics[width=7cm,angle=-0]{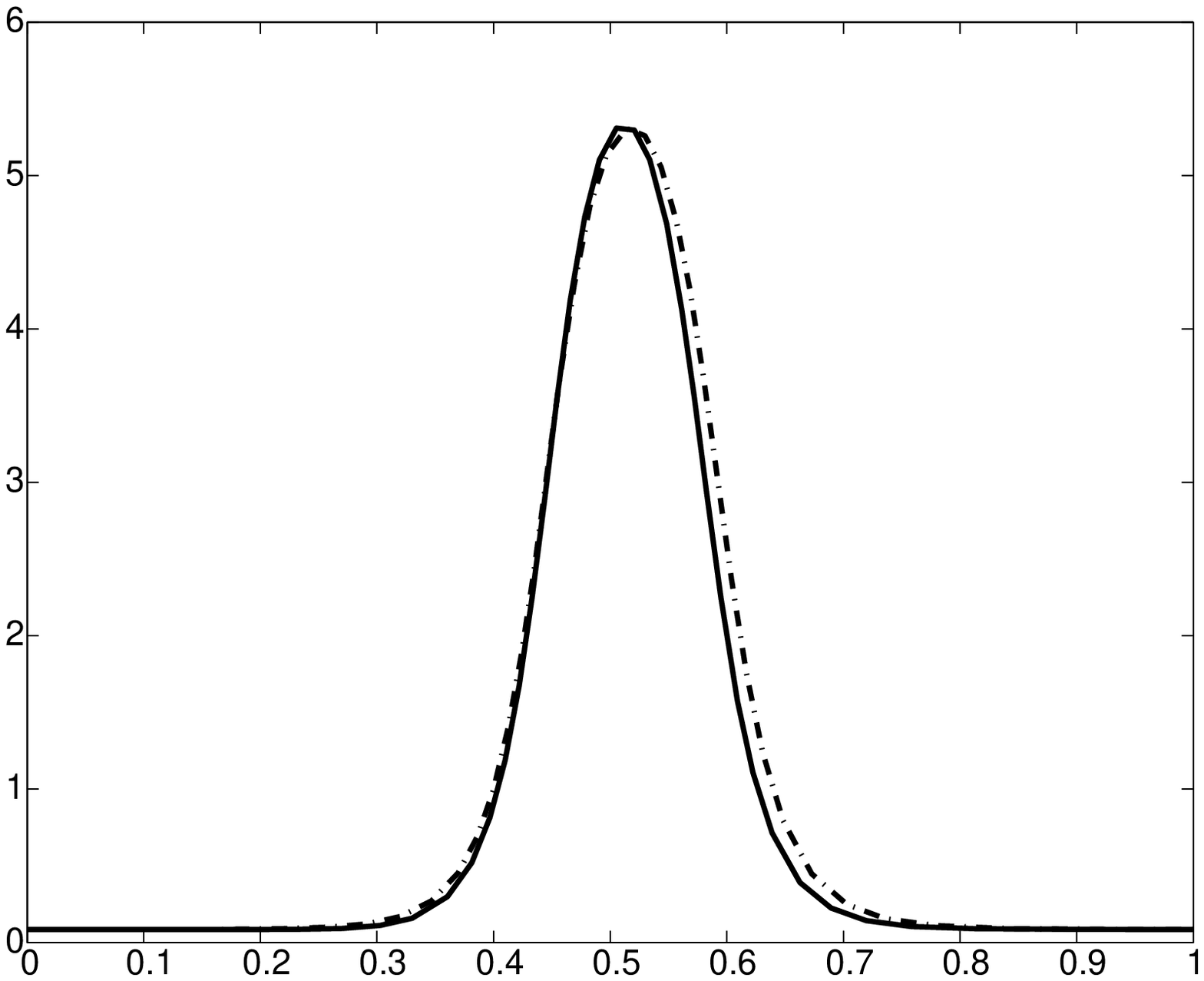}}
\end{center}
\caption{Two pairs of solution to (\ref{sjj}) for $J=0.5$ and
$\gamma=0.17$ that lie on the left and right branch of the 'back-bending' curve (see Fig. \ref{ivj}(b)) are compared. Solutions on the right branch are given in dashed-dotted lines.}
\label{solivjlain}
\end{figure}

If we keep increasing the applied bias current, the velocity of
the fluxon increases up to the maximum velocity
$|u_{max}^{[1|0]}|$. This maximum velocity is a function of
$J,\,S,\, \alpha$ \cite{gwu}. By now, the analytic form of the
maximum velocity for this state is still unknown.

Another branch is found when we use one solution on the right
branch of the IV-characteristic for $J>1$ as the initial solution
of AUTO. Using this initial condition, then we decrease the
parameter $J$ up to $J=0.5$. In fact, we get a backbending curve
for the IV-characteristic of the case $J<1$. The curve is given in
Fig. \ref{ivj}(b). If for $J>1$, we see hardly any changes when
comparing solutions in the first junction, while something
different happens in the second junction (Fig. \ref{solbackban}),
for the case we have the opposite result. In junction 2 we almost
have seen nothing, while there is a difference in the solutions in
the first junction. In Fig. \ref{solivjlain} we compare two
solutions on the left and right branch for the same value of
$\gamma$.

A further analysis is needed to see the stability of solutions
lying on this branch. We conjecture that all the solutions are
unstable. The same argument might be used as in the case $J>1$
before. Numerical simulation confirms as well that the solutions
go to the corresponding solutions at the other branch. We have
done that also using PDE-solver where we use a solution resulted
by AUTO as the initial solution of the PDE-solver.

\section{Conclusion}
\label{sec4}

We have given several calculations showing that stacked-Josephson
junctions have very rich behavior. Just with $[1|0]$ state and
changing $J$, we already have qualitatively different
IV-characteristics, while there are many more states that can
exist in the junctions.

How the IV-characteristics corresponding to the state $[n|m]$
change when $J$ passes through $1$ needs to be investigated
further. Also the stability of the various patterns needs to be
determined.

Goldobin {\it et al.} \cite{gwu} state that for the $[1|0]$ state
not only the condition $J<1$ leads to $|c|>|c_+|$, but also the
asymmetry of the applied bias current of the two-fold stack, i.e.
when we have $\gamma^1>\gamma^2$. Therefore, it might be
interesting to consider also in the near future the possibility of
having branches of the system under condition
$\gamma^1\neq\gamma^2$.

\end{document}